\documentclass{ws-ijmpd}

\usepackage{acronym}
\usepackage[usenames]{color}
\usepackage[bookmarksnumbered, bookmarksopen, breaklinks, colorlinks]{hyperref}

\newcommand{\AEI}{Albert-Einstein-Institut, Max-Planck-Institut f\"ur
Gravitationsphysik\\ D-30167 Hannover, Germany}
\newcommand{\Leibniz}{Leibniz Universit\"at Hannover\\ D-30167 Hannover, Germany}

\newcommand{\mat}[1]{{\bf #1}}
\newcommand{\T}{{\mathrm{T}}}
\newcommand{\abs}[1]{\left\lvert #1 \right\rvert}
\begin{document}


\markboth{Drew~Keppel}
{Use of Singular-Value Decomposition in Gravitational-Wave Data Analysis}

\title{Use of Singular-Value Decomposition in Gravitational-Wave Data Analysis}

\author{Drew~Keppel}  

\address{
\AEI\\
\Leibniz\\
drew.keppel@ligo.org
}

\maketitle
\acrodef{CBC}{compact binary coalescence}
\acrodef{GW}{gravitational-wave}
\acrodef{MCMC}{Markov chain Monte Carlo}
\acrodef{PN}{post-Newtonian}
\acrodef{SN}{supernova}
\acrodefplural{SN}[SN]{supernovae}
\acrodef{SNR}{signal-to-noise ratio}
\acrodef{SVD}{singular value decomposition}

\begin{abstract}
Singular-value decomposition is a powerful technique that has been used in the
analysis of matrices in many fields. In this paper, we summarize how it has
been applied to the analysis of gravitational-wave data. These include
producing basis waveforms for matched filtering, decreasing the computational
cost of searching for many waveforms, improving parameter estimation, and
providing a method of waveform interpolation.
\end{abstract}

\section{Introduction}

Within the field of \ac{GW} data analysis, there are several difficult problems
that need to be addressed in order to extract information from the data that
will be taken by advanced \ac{GW} detectors.  These include determining when a
\ac{GW} signal occurred, where this signal came from, and what the properties
of the signal's source are. In this paper, we summarize how \ac{SVD} has been
used to ease the computational burden of answering these questions.

This paper is outlined as follows: Sect.~\ref{sect-SVD} gives a brief overview
of \ac{SVD}, Sect.~\ref{sect-waveforms} describes how \ac{SVD} can be used to
approximate different classes of waveforms, Sect.~\ref{sect-compdetstat}
illustrates how \ac{SVD} has be used to construct composite detection
statistics, Sect.~\ref{sect-waveinterp} discusses a method of waveform
interpolation based on \ac{SVD}, and Sect.~\ref{sect-paramest} summarizes the
use of \ac{SVD} in different aspects of parameter estimation.

\section{Singular Value Decomposition}
\label{sect-SVD}

\ac{SVD} decomposes an arbitrary matrix $\mat{H}$ into three components:
\begin{equation}
\mat{H} = \mat{V} \mat{\Sigma} \mat{U},
\end{equation}
where $\mat{U}$ is a unitary matrix of basis vectors such that the columns of
$\mat{U}^*$ are the right eigenvectors of $\mat{H}^* \mat{H}$, $\mat{V}$ is a
unitary matrix of reconstruction coefficients such that the columns of
$\mat{V}$ are the left eigenvectors of $\mat{H} \mat{H}^*$, and $\mat{\Sigma}$
is a diagonal matrix of singular values whose non-zero components are the
square root of the eigenvalues associated with $\mat{H}^* \mat{H}$ or $\mat{H}
\mat{H}^*$.\footnote{Throughout this paper, $\mat{M}^*$ denotes the complex
conjugate-transpose of matrix $\mat{M}$.} The eigenvectors of $\mat{U}$ and
$\mat{V}$ are ordered such that the entries of $\mat{\Sigma}$ are in descending
order.

We will see in the following sections that the properties of $\mat{U}$,
$\mat{V}$, and $\mat{\Sigma}$ have been used in different contexts.

\section{Classes of Waveforms}
\label{sect-waveforms}

\ac{GW} signal waveforms can be classified into two groups: those that are well
modelled, and those that are not.\footnote{One reason a waveform may not be
able to be well modelled is because of complex and possibly chaotic physical
processes at work in the production of the \ac{GW} signal (\emph{e.g.}, the
presence of matter in a \ac{GW} generating system).} Collections of waveforms
of both types can be called \emph{waveform catalogs}, however here we restrict
use of that term for non-well-modelled waveforms. We will refer to waveforms
that are well modelled as \emph{template banks}.  Further discussion of
temaplte banks and waveform catalogs can be found in Sect.~\ref{subsect-banks}
and \ref{subsect-catalogs}, respectively.

\subsection{Waveform Template Banks}
\label{subsect-banks}

An example of a type of waveform that can be used to construct a template bank
is a \ac{GW} signal from the inspiral of a \ac{CBC}. These waveforms are well
described theoretically by the \ac{PN} approximation. This has allowed the
construction of a metric on the signal parameter space that characterizes the
distance between two points on parameter space based on the mismatch between
the two waveforms.\cite{Owen1996} Using this metric, a template bank can be
constructed with these waveforms such that any point within a chosen region of
parameter space will be no more than a specified distance from the nearest
template, commonly chosen such that maximum mismatch between any template in
the bank and any signal in the region of interest is $3\%$.

In order to meet this requirement with the fewest number of templates, for a
two dimensional parameter space with a constant metric, adjacent templates will
have a mismatch of 9\%, which means they have an overlap of 91\%.  Cannon
\emph{et al.} investigated how the \ac{SVD} could compress a matrix $\mat{H}$
whose rows are composed of the highly overlapping waveform time-series from
such a template bank.\cite{Cannon2010} They derived the expected fractional
\ac{SNR} loss $\delta\rho / \rho$ of approximating the waveforms in the
template bank by the truncation of the \ac{SVD} of $\mat{H}$ to be
\begin{equation}
\frac{\delta\rho}{\rho} = \frac{1}{2N} \sum_{i=N'+1}^{N} \sigma_i^2,
\end{equation}
where $\sigma_i$ is the $i$th entry of $\mat{\Sigma}$, $N$ is the number of
rows of $\mat{H}$  or twice the number of templates in the template bank, and
$N'$ is where the \ac{SVD} is truncated.  In addition, the maximum
phase-averaged fractional \ac{SNR} loss can be bounded by
\begin{equation}
\left(\frac{\delta\rho}{\rho}\right)_\mathrm{max} < \frac{1}{2}
\sigma_{N'+1}^2,
\end{equation}
which can be derived from the unitarity of the reconstruction coefficients.

This approach to waveform compression has been implemented within a low-latency
\ac{GW} search for inspiral waveforms as one of the techniques to reduce the
computational cost of data processing.\cite{lloid2011} Similar techniques have
also been applied to other types of \ac{GW} signals using Gram-Schmidt
orthogonalization.\cite{redbasis2011a,redbasis2011b}

\subsection{Waveform Catalogs}
\label{subsect-catalogs}

Waveform catalogs are necessary for waveforms that are not well modelled
theoretically. Such waveforms can arise from numerical simulations that are
either too computationally expensive to study the parameter space thoroughly
(\emph{e.g.}, numerical simulations of merging spinning binary black holes), or
from numerical simulations that include matter (\emph{e.g.}, numerical
simulations of \acp{SN} or coalescing binary neutron stars).

Ref.~\refcite{bradyraymajumder2004} first proposed decomposing waveform
catalogs of \ac{SN} waveforms using Gram-Schmidt orthogonalization.
Ref.~\refcite{heng2008} expanded on this approach using the \ac{SVD} and
comparing it to Gram-Schmidt orthogonalization.  Both approaches were found to
perform similarly, being able to extract and prioritize waveform
characteristics from the \ac{SN} waveform catalogs. However, in the \ac{SVD}
only a single decomposition of the catalog was needed, and additional
information was obtained. Namely, features that were present in many waveforms
produced basis vectors with large singular values, whereas features present in
only a few produce basis vectors with small singular values.

\section{Detection Statistics}
\label{sect-compdetstat}

Constructing detection statistics when searching for \ac{GW} signals is
different depending on what type of search you are doing. Searches can be
classified into two groups, those searching for known signal waveforms and
those searching for unknown signal waveforms.

The standard approach to search for any known signal within \ac{GW} data from a
collection of \ac{GW} waveforms is to matched filter the data, producing a
time-series of \ac{SNR}, with each of the template waveforms. \ac{GW} signals
are then identified by identifying which templates at certain times produce
large values of \ac{SNR}.\cite{wainstein}

If one is only interested in answering the time component of the question, a
\emph{composite detection statistic} can be used, which tells whether the data
looks like anything from a collection of template waveforms.\cite{wainstein}
Cannon \emph{et al.}~showed how the \ac{SVD} can be used to construct such a
composite detection statistic $\Gamma$ associated with a region of parameter
space covered by a template bank of the form
\begin{equation}
\Gamma = \sum_{k=1}^{N'} \frac{\sigma_k^2}{\sigma_k^2 + 2N / \langle A^2
\rangle} \rho_k^2,
\end{equation}
where $\langle A^2 \rangle$ is the expected signal amplitude squared and
$\rho_k$ is the \ac{SNR} associated with the $k$th basis
vector.\cite{Cannon2011cds} This statistic was derived using the orthonormality
of the reconstruction coefficients and the basis vectors, as well as the
ordered significance given by the singular values.  It shows promise in
reducing the computational cost of searching \ac{GW} data for signals of known
waveforms.

When searching for a signal of unknown waveform, the search can be performed by
looking at other aspects of how the signal interacts with the detectors. In
particular, when there is a network of \ac{GW} detectors in operation, the data
from these detectors can be combined in such a way to produce coherent data
streams that are direction dependent.\cite{gurseltinto89,flanhughes97}

Wen outlined how the detector responses associated with a given direction can
be coherently combined and condensed using the \ac{SVD}.\cite{wen2008} The
\ac{SVD} of the network detector responses $\mat{A}$ provides the basis vectors
$\mat{U}$, singular values $\mat{\Sigma}$,  and reconstruction coefficients
$\mat{V}$.  The data from the different detectors $\mat{d}$ can be recombined
into new data streams $\mat{d}'$ using the reconstruction coefficients
\begin{equation}
\mat{d}' = \mat{V}^{-1} \mat{d}.
\end{equation}
The detection statistics for an unmodelled \ac{GW} search could then be
constructed using
\begin{equation}
\Gamma = \sum_{k} w_k \abs{d_k'}^2,
\end{equation}
where $w_k$ is a chosen weight associated with different basis vectors, and the
sum can be over different combinations of the new data streams. Different
choices for each of these parameters results in different optimizations of the
detection statistic.

\section{Waveform Interpolation}
\label{sect-waveinterp}

Producing the waveforms in the first place can be a computational challenge,
which is the case for binary black hole coalescence waveforms produced by
numerical relativity. To help solve this problem, there have been several
efforts to parameterize these waveforms, allowing waveforms from arbitrary
points in parameter space to be produced.\footnote{References for these efforts
can be found in Ref.~\refcite{Cannon2011interp}.} These efforts are based on
models of the waveforms that need to be tuned. Cannon \emph{et al.} have
pursued a different approach by showing that the basis vectors from the
\ac{SVD} of waveforms from a template bank approximately enclose the space of
waveforms from that region of parameter space.\cite{Cannon2011manifold}
Further, they have also laid out a procedure that could be used to interpolate
waveforms to high accuracy from arbitrary points in parameter space using the
components of the \ac{SVD}.\cite{Cannon2011interp} This is done by
interpolating the reconstruction coefficients. Once the interpolated
reconstruction coefficients $\mat{V}'$ are in hand, the interpolated waveforms
can be generated using
\begin{equation}
\mat{H}' = \mat{V}' \mat{\Sigma} \mat{U},
\end{equation}
where $\mat{H}'$ is a matrix containing interpolated waveforms at the locations
where $\mat{V}'$ was interpolated. With a sufficient density of numerical
relativity waveforms, highly accurate \ac{SVD} interpolated waveforms could be
produced for arbitrary points in parameter space.

\section{Parameter Estimation}
\label{sect-paramest}

Once it is known that there exists a \ac{GW} signal in a data set, the next
question that comes up is what are the parameters of the source of the signal.
Several sets of authors have used the \ac{SVD} toward this goal in different
contexts.

Wen identified how the \ac{SVD} could be used to construct \emph{null data
streams} and \emph{semi-null data streams}.\cite{wen2008} Null data streams are
linear combinations of data from a network of \ac{GW} detectors that will
contain no \ac{GW} signals,\cite{wenschutz2005} which are associated with
zero-valued singular values from the \ac{SVD} of the response of a network of
\ac{GW} detectors. Similarly, semi-null data streams are associated with
small-valued singular values from the \ac{SVD} of the response of a network of
\ac{GW} detectors. Wen \emph{et al.} showed how analyzing these semi-null data
streams could improve the accuracy with which one could locate the source of a
\ac{GW} signal on the sky.\cite{wenfanchen2008}

Wen also explained how the \ac{SVD} could be used to extract a \ac{GW} signal
waveform from the data of a network of \ac{GW} detectors.\cite{wen2008} This
was done by truncating the \ac{SVD} of the detector response matrix in order to
regularize it. The extracted waveform $\mat{h}^\T$ is then the data from all
detectors $\mat{d}$ combined using the inverse of the regularized detectors
response matrix,
\begin{equation}
\mat{h}^\T = \left( \mat{A}^\T \right)^{-1} \mat{d},
\end{equation}
where $A^\T_{ij} = \sum_{k=1}^{N'} v_{ik} \sigma_k u_{kj}$ is the detectors'
response matrix regularized by truncating the \ac{SVD} after the $N'$th
singular value.

R\"{o}ver \emph{et al.} applied the \ac{SVD} of waveform catalogs to
the parameter estimation problem.\cite{rover2009} Using the basis vectors
associated with the \ac{SVD} of a \ac{SN} waveform catalog, a parameter
estimation search based on \ac{MCMC} methods extracted the amplitudes of the
signal associated with each of the basis vectors. By associating the extracted
amplitudes with those of the waveforms in the catalog, physical parameters of
the simulated signal could be estimated.

Extending the previously mentioned work, Logue \emph{et al.} applied the
\ac{SVD} to multiple waveform catalogs, each associated with different \ac{SN}
production mechanisms or different types of stellar cores.\cite{logue2012} By
computing the Bayes odd ratios associated with these different models by nested
sampling techniques, the correct model for the \ac{SN} can be determined.

Finally, Cannon \emph{et al.} discussed how the likelihood function, which is
the quantity computed for parameter estimation, could be interpolated to
decrease the computation cost of parameter estimation
searches.\cite{Cannon2011interp} Using the interpolated reconstruction
coefficients, the interpolated likelihood function would be given by
\begin{equation}
\ln \left( \mat{\Lambda}' \right) = \mat{V}' \mat{\rho}^2,
\end{equation}
where the locations the likelihood function is estimated are given by the
locations where the reconstruction coefficients are interpolated.

\section{Conclusion}

\ac{SVD} has been applied with good results in a number of contexts within
\ac{GW} data analysis. From waveform compression and interpolation, to deriving
new detection statistics, to improving parameter estimation, the powerful
properties of the \ac{SVD} have proved to be of great use.

\section*{Acknowledgments}

We would like to thank Kipp Cannon and Chad Hanna for the countless fruitful
discussions we have had on this topic, and Bruce Allen and Ray Frey for their
useful comments on this manuscript. We would also like David Blair and Linqing
Wen for extending to us the invitation to attend the Third Galileo - Xu Guangqi
Meeting. This work was supported from the Max Planck Gesellschaft and has LIGO
document number {LIGO-P1100207}.

\end{document}